# Measuring the Correlation of Personal Identity Documents in Structured Format

Sachithra Dangalla[+], Chanaka Lakmal[*], Chamin Wickramarathna[¶], Chandu Herath[§], Gihan Dias[ε], Shantha Fernando[ǂ]
*Department of Computer Science and Engineering, University of Moratuwa*
Sri Lanka
[+]sachithradangalla@gmail.com, [*]ldclakmal@gmail.com, [¶]chaminbw@gmail.com, [§]cbherath93@gmail.com, [ε]gihan@uom.lk, [ǂ]shantha@cse.mrt.ac.lk

***Abstract*—Personal identity documents play a major role in every citizen's life and the authorities responsible for validating them typically require human intervention to manually cross-check multiple documents belonging to an individual. The world is rapidly replacing physical documents with digital documents where every piece of data is stored digitally in a machine-readable and structured format. In this paper, we describe a technique to extract identity data from a structured data format and calculate a normalized correlation score for personal identity documents. Experimental results show that the proposed technique effectively calculates the correlation score for personal identity documents.**

## I. INTRODUCTION

Personal identity is a dataset that can be used to recognize a person and differentiate him from others and an identity document is a piece of documentation that is specifically designed to prove the identity of an individual [1]. Personal identity documents play a major role in identity validation processes that are typically carried out in government or private institutions. One popular use case of personal identity documents is their use in validating a set of documents belonging to a single citizen where the institutions usually require human intervention and knowledge to intelligently analyze and identify whether they belong to the same person or not. In this paper, we introduce a technique to extract identity data based on the attribute names and calculate a normalized real-time correlation score for a set of personal identity documents corresponding to five super identity attributes we have identified.

With the world drifting away from the use of physical documents, digital documents are taking over the world and replacing information with digital machine-readable data. These machine-readable data are typically in structured data format where data are stored in the key-value form. We propose our technique for data in structured format which eliminates the requirement to explicitly retrieve values corresponding to a specific target attribute from raw data.

This paper is structured as follows. Section II discusses the literature of existing mechanisms that were reviewed and selected and not selected for this solution with justifications. Section III describes the methodology we followed to define the algorithm we propose along with the experimental results of the technique. The proposed algorithm is discussed in Section IV. Section V discusses the future enhancements that could fine-tune this technique.

## II. LITERATURE SURVEY

Personal identity is a dataset that can be used to recognize a person and differentiate him from others. Clarke defines identity as "a presentation or role of some underlying entity" [2]. Miller in his book has discussed 3 approaches to prove a person's identity [3].

1. Something you have – The availability of a physical object in possession of the person. (Eg. key)
2. Something you know – A predefined fact or knowledge that is known by the person. (Eg. password)
3. Something you are – Biometrics or measurable personal traits (Eg. fingerprint)

Woodward, Horn, Gatune and Thomas state that this personal identity is the key to provide the right privileges to the right person with right access at the right time [4].

An identity document is a piece of documentation that is specifically designed to prove the identity of an individual [1]. It belongs to the first of the three categories introduced by Miller [3]. The term legal identity, usually assigned to every citizen, is referring to the fact that all human beings should be known and individualized by their registry office [5]. The legal process of issuing identity documents may differ according to the country or the purpose of the issued document but the underlying process of issuing an identity document has 4 common steps [6], [7]:

1. The applicant registers the document request with the municipal authorities.
2. The applicant proves his identity with supportive documents and/or biometrics.
3. The authorities collect, and fill required information and process the document.
4. Authorities accept/deny the issuance of the document.

Lisbach discusses in detail about analyzing the identity of people. It classifies attributes into 4 categories as space-related attributes that include address, place of birth and nationality, time-related attributes that include date of birth, classifying attributes that include gender and identification codes that include tax reference, passport number or social security number [8]. We have directly used these four categories to determine the identity attributes in our correlation score calculation.

While there has been a great deal of work on person name disambiguation, our problem stands out to be unique since we use a static approach with real time data processing.

Fleischman and Hovy [9] have presented an algorithm using *Maximum Entropy Model* to find whether the same name in multiple documents refer to the same person or not. They have used a two-step approach where they first identify concept-instance pairs in documents and then use probabilistic measures and agglomerative clustering techniques to cluster similar pairs together. Niu, Li and Srihari [10] have addressed the cross-document person name disambiguation using a supervised technique that focuses on the number of occurrences of a name in a document set. In certain countries such as Sri Lanka and India, person names are lengthy and can be subjected to minor changes upon translation from Sinhala or Tamil to English, whereas the above approaches require exactly matched short person names to be matched.

Torvik, Weeber, Swanson and Smalheiser [11] discuss about name disambiguation with regard to publications on Medline, a bioinformatics research database. The paper discusses about name attributes which were used to determine the possible attributes that can be categorized under the name of a person. It uses medicine-related attributes with weights for calculation and we use a similar approach for attributes related to personal information.

Euzenat and Valtchev [12] discuss about similarity measuring in web ontology language. They define a similarity measuring algorithm based on relationships among the entities and a class structure. We used the pair-wise local similarity measurement used in this research in our approach where we determine the pair-wise similarity between documents. However, when averaging the pair-wise similarity, they have considered the maximum count as the denominator, but we consider the total number of unique items in both elements in the pair as our denominator.

Correlation Preserving Index (CPI) can be used to discover structures in high dimensional document space. In their research paper, Nagaraj and Thiagarasu [13] discuss how Ridge Regression with Eigen values is used for document clustering in order to measure similarity. The algorithm is comprehensive and is fine-tuned to be used with high-dimensional document space whereas our approach focuses on structured data that predefines the attributes of the data to be considered for measuring similarity.

When measuring the name similarity, the translation between languages and the differences in the alphabets of the languages cause variations in representations of the names. In order to address this problem, we use phonetic similarity in the name similarity measure.

Soundex is a coding system that allows to transform a name into a code according to the way it sounds rather than the way it is spelled [14]. Many researches have made improvements on the traditional Soundex implementations to increase the precision and the accuracy of the output code with the use of N-gram substitutions [15] and approximate string matching techniques [16]. Weiling, Margaretha and Nerbonne have introduced an approach to iterate alignment and information-theoretic distance assignment until both remain stable, and then evaluate the quality of the resulting phonetic distances by comparing them to acoustic vowel distances [17].

Kondrak has introduced a phonetic similarity and alignment measure by using an extended set of edit operations, local and semiglobal modes of alignment, and the capability of retrieving a set of near-optimal alignments [18]. His approach focuses on preserving the salience of the words whereas our requirement is to identify name segments that refer to the same name by considering the phonetic similarity. In his thesis, Heeringa discusses the application of Levenshtein distance to transcriptions of word pronunciations [19]. In our approach, we directly make use of the improved Soundex in the name similarity measure.

In order to measure the textual similarity between the document content, we analyzed existing string similarity measures and techniques. A large number of string similarity measures exist that calculates the similarity based on edit distance and longest common subsequence [20], [21]. We conducted an experiment to select the best string similarity measure that provides the optimum results.

Data extraction from paper documents or unstructured data into a structured data format can be achieved using data extraction mechanisms [22].

## III. METHODOLOGY

Correlation between structured data can be measured by considering the similarity of the values of similar attributes. This measure can be too vague when the domain or the context of the data can vary immensely. However, if this domain or context is limited to personal identity data, the correlation measure can be fine-tuned to measure and calculate a score to define the correlation between these data to a single personal identity. The focus of this research is to identify personal identity attributes, experiment on how these attribute values can vary for a single personal identity and then use the results obtained to construct the algorithm to calculate the correlation score.

### *A. Super Identity Attributes*

Identity characteristics other than the name of a person, can be grouped into classes of attributes that share similar features as follows [8].

- Space-related attributes: Address, place of birth and nationality
- Time-related attributes: Date of birth
- Classifying attributes: Gender
- Identification codes: Tax reference, passport number or social security number

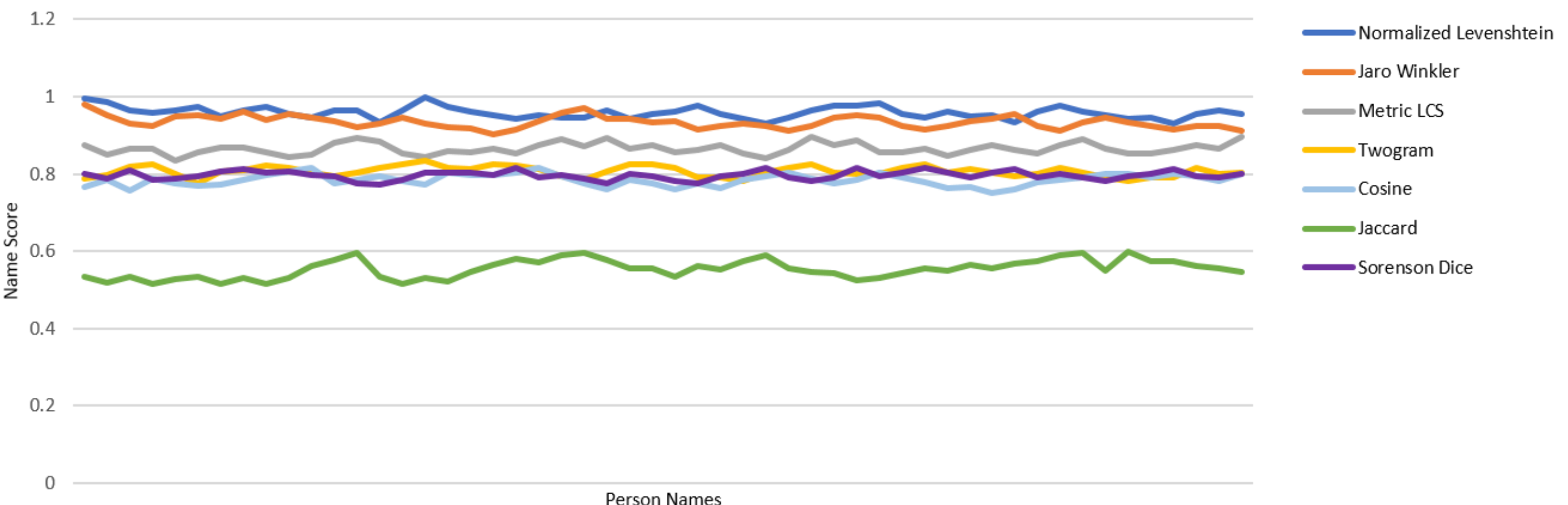

Fig. 1. Results of string similarity methods used in the algorithm

Furthermore, identity characteristics can also be categorized into two classes:

- Attributes that define past incidents: Date of birth, place of birth
- Attributes that define current status: Current address

By considering both the above approaches, we selected five super identity attributes that are common in a majority of personal identity documents.

1. Name
2. Date of birth
3. Gender
4. Address
5. National Identification Code [1]

### B. *Value extraction and correlation score of attributes*

Value extraction refers to the process that identifies the values relevant to the five super attributes from structured data format. This value extraction process is based on comparing the attribute name with a predefined set of attribute names and classifying under one of the five super attributes. The algorithm calculates a score for each super attribute in each document and then use the average to calculate the correlation score for each document.

### C. *Correlation score of document*

The calculated correlation scores of each super identity attribute of each document are combined to calculate the correlation score of a single document from a set of documents. A weighted approach is not suitable for this calculation because all five of the super identity attributes contribute equally to define the identity of a person. Therefore, the final correlation score is the average of the scores of all attributes.

[1] – The National Identification Code refers to a code or number issued uniquely to a citizen by the government. The experiment to test the results was carried out with data of Sri Lankan citizens and the NIC (National Identity Card number) was considered as the fifth identity attribute.

### D. *Experimental results*

Two experiments were carried out to test the distribution of the attribute scores in the correlation score algorithm by collecting information of five identity documents from Sri Lankan citizens. The documents were the National Identity card, Passport, Birth certificate, Driving License and Marriage certificate.

The first experiment was conducted to test the strength of Levenshtein similarity in measuring the string similarity. The string similarity method in the name score was subjected to the following normalized string similarity measuring methods.

1. Cosine Similarity [23]
2. Jaccard Similarity [23]
3. Jaro Winkler [24]
4. Metric LCS [25]
5. Normalized Levenshtein [19]
6. Sorensen Dice [20]
7. Two-gram similarity [20]

The results proved that Levenshtein is the optimal similarity measuring algorithm to measure the similarity of names adhering to Heeringa's theory [19]. The results of the first experiment are shown in Fig. 1.

The second experiment was a survey where the deviations of the attribute scores were observed. The results of the survey experiment are shown in Fig. 2.

The name score displayed the highest deviation which was caused by the differences in representation of the names in different documents. Gender score displayed the least deviation where every document in each test case represented the same gender class.

## IV. ALGORITHM

### A. *Name*

In the experiment we conducted, in the structured data format, the values have nested forms and have multiple representations of a single attribute, especially of the name.

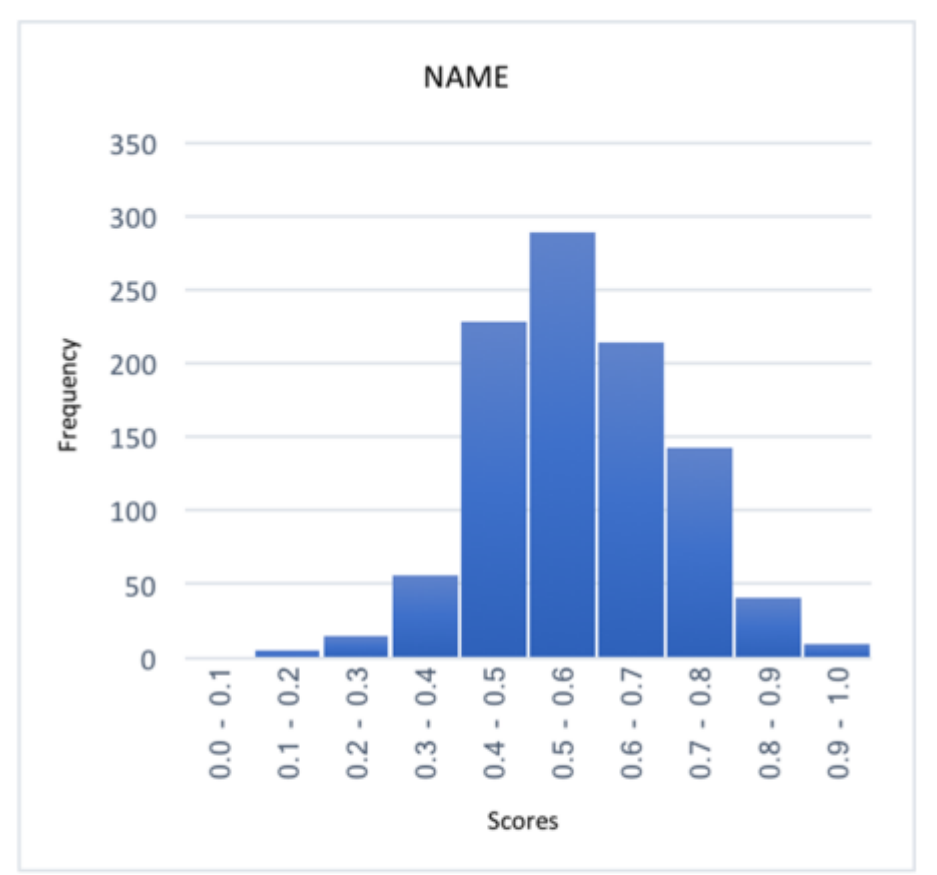

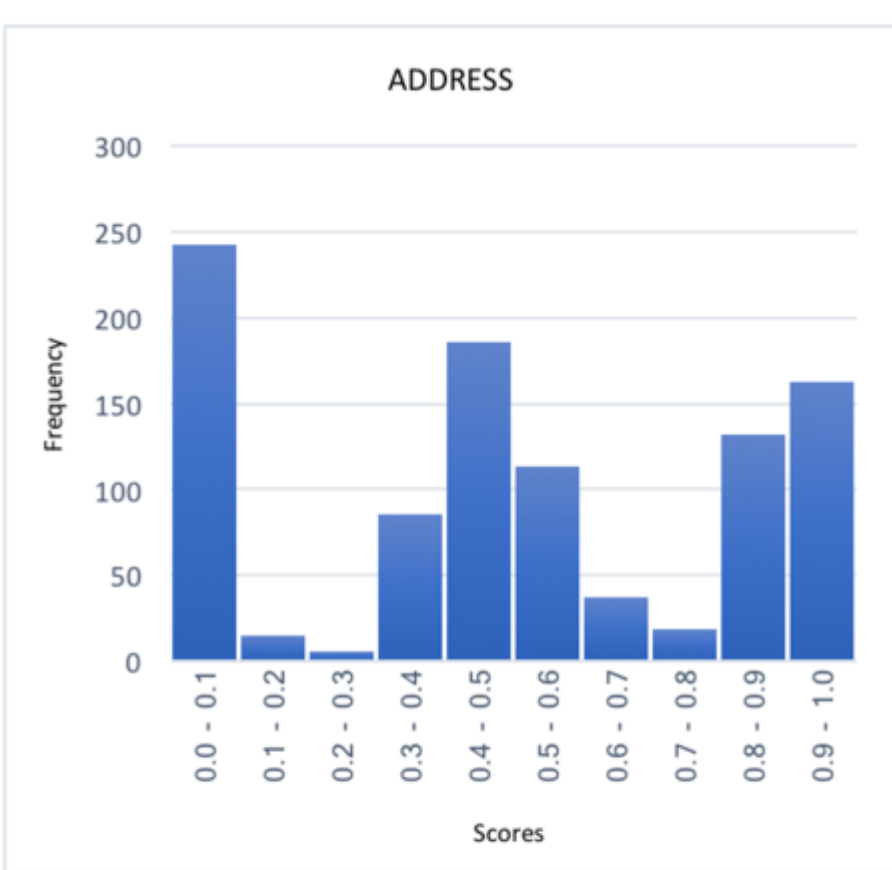

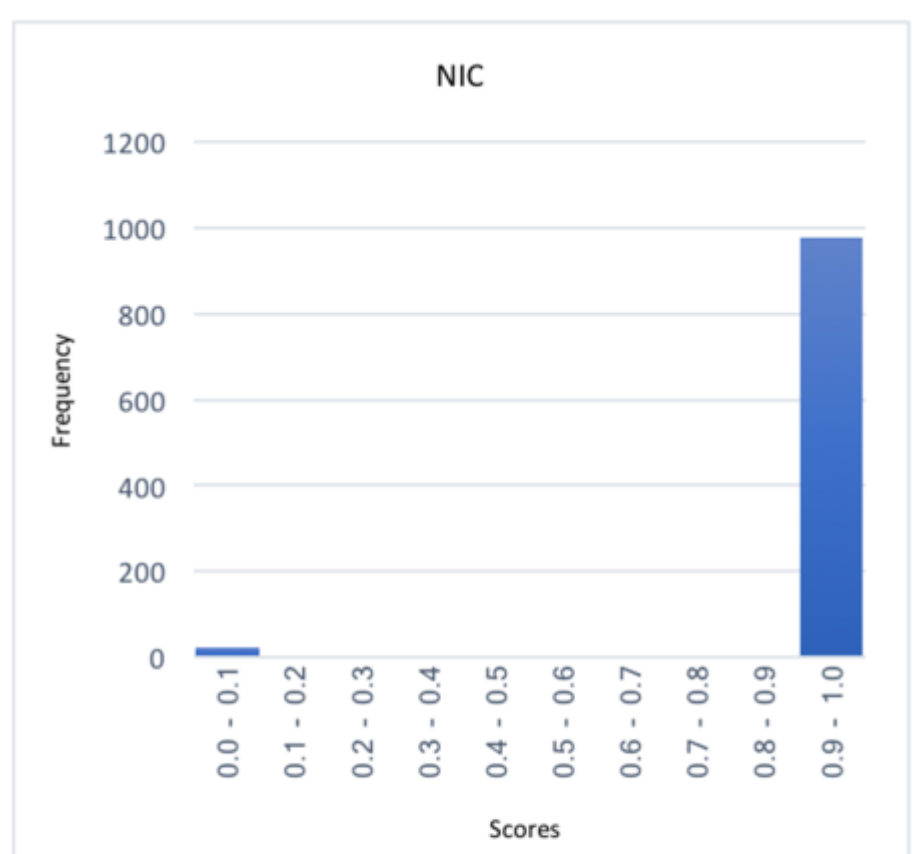

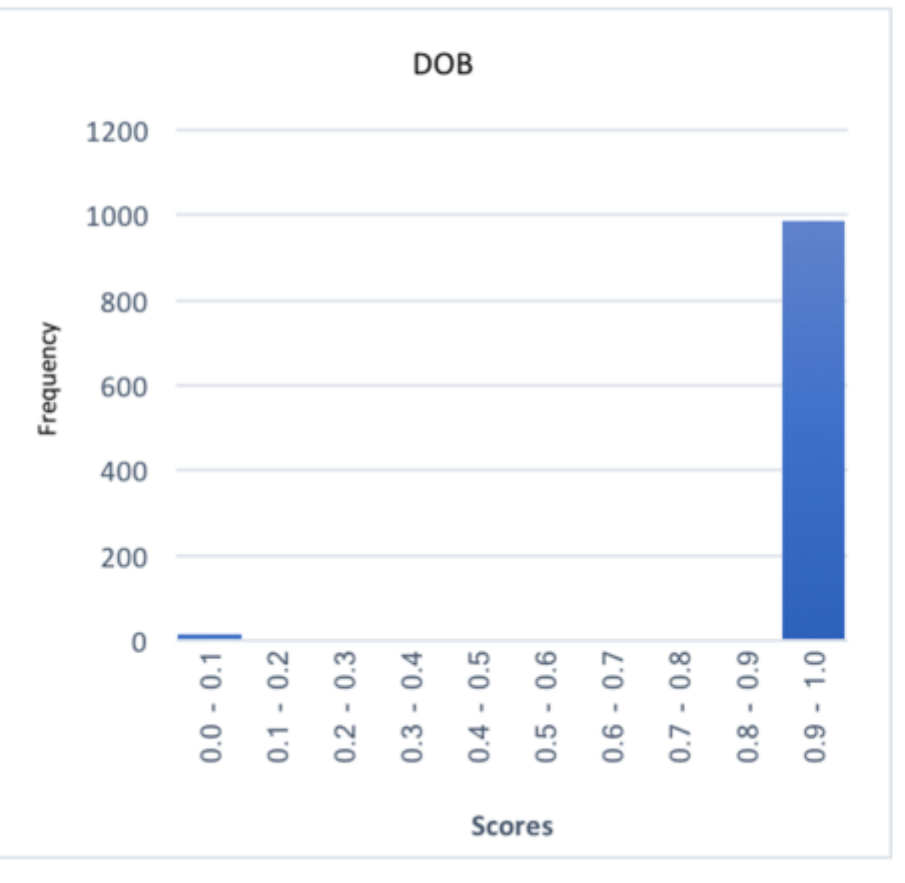

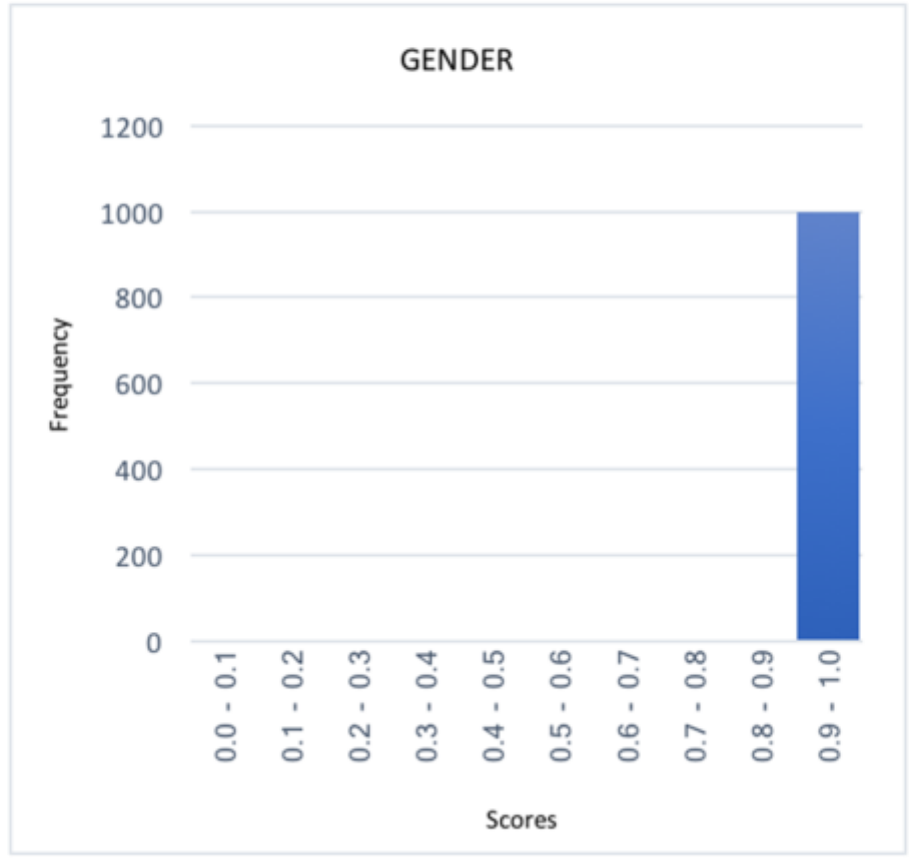

Fig. 2. Frequency distribution of attribute scores

The name of the attribute and the nested structure solely depend on the document structure. After analyzing a set of 20 personal documents, we identified the base attribute names and the possible ordering of the attributes for the name as follows [2].

It was observed in the experiment that the name segments and the order of the name segments represented in the documents vary with each document structure and the optimum results were not obtained when the name segments were concatenated in the given order of the document [3]. Therefore, we came up with the order of name segments as presented in the attribute set which was used to concatenate the name segments. If a document contained an attribute that is not available in our attribute set, the concatenation was done according to the order presented in the document.

The attribute set = {<initials>, <first_name>, <middle_name>, <name>, <full_name>, <surname>, <other_names>, <last_name>}

---

[2] – The analyzed documents are passport, national identity card, birth certificate, driving license, marriage certificate, the university ID card, motor vehicle registration certificate, vehicle insurance card, mortgage deed, university transcript, university degree certificate, will (testament), bank statement, credit card statement, divorce papers, income tax sheets.

[3] – Eg. Passport contains <surname> <other_names> and personal medical record contains <last_name> <first_name> which gives different values for the same name when concatenated in the order of attributes presented in the document.

Steps for measuring the correlation between the name values:

1. Obtain a single value by concatenating the values in the document according to the order of the attributes in the attribute set.
2. Convert each name segment to its phonetic representation using Soundex [14].
3. For each pair of documents $d_i$ and $d_j$
   - 3.1. If <initials> is available in one of the two documents, compare the <initials> with the first letters of the names of the other document
   - 3.2. Identify overlapping name segments by comparing the phonetic representations
   - 3.3. Calculate Order Similarity ($OS_{d_i,d_j,name}$) by calculating the Jaccard Coefficient [23].

$$OS_{d_i,d_j,name} = \frac{\text{Overlapping segments in the same order}}{\text{Unique name segments in } d_i \text{ and } d_j} \qquad (1)$$

   - 3.4. Calculate String Similarity $SS_{d_i,d_j,name}$ of the name segments with similar phonetic codes using Levenshtein distance

$$SS_{d_i,d_j,name} = \frac{\sum \text{Levenshtein similarity}}{\text{Unique name segments in } d_i \text{ and } d_j} \qquad (2)$$

   - 3.5. Calculate the pair-wise correlation score of name

$$\left(\mathrm{CS}_{\mathrm{d_i,d_j,name}}\right) = \frac{\mathrm{OS}_{\mathrm{d_i,d_j,name}}+\mathrm{SS}_{\mathrm{d_i,d_j,name}}}{2} \tag{3}$$

4. Calculate the correlation score for name for $\mathrm{d_k}$

$$CS_{d_k,name} = \left(\frac{1}{n-1}\right)\sum_{\substack{i=0 \\ i\neq k}}^{n} CS_{d_k,d_i,name} \tag{4}$$

where $\mathrm{d_k}$ is the k-th document and $n$ is the number of documents that includes <name>

*B. Date of Birth*

The attribute set = {<date_of_birth>, <dob>, <birth_date>}

Child attribute set = {<date>, <day>, <d>, <month>, <m>, <year>, <y>}

The date of birth was analyzed by checking the exact matching values for the date, month and year.

If different values were obtained, a candidate value $(\mathrm{Candidate_{dob}})$ will be selected based on majority voting and the score will be calculated against the candidate.

$$\mathrm{CS}_{\mathrm{d_k,dob}} = \begin{cases} 1, & \mathrm{DOB}_{\mathrm{d_k}} = \mathrm{Candidate_{dob}} \\ 0, & \text{otherwise} \end{cases} \tag{5}$$

*C. Gender*

The attribute space = {<gender>, <sex>}

For the scope of this paper, we assumed only two gender values (male and female).

The similarity was measured by classifying the value into one of the two target values and then checking if the values are of the same class. The classification process was based on checking textual similarity with a target set of values.

Target value set for class 1 = {<f>, <female>}

Target value set for class 2 = {<m>, <male>}

If different values were obtained, the candidate method was used as in Section IV Part B.

$$\mathrm{CS}_{\mathrm{d_k,gender}} = \begin{cases} 1, & \mathrm{GENDER}_{\mathrm{d_k}} = \mathrm{Candidate_{gender}} \\ 0, & \text{otherwise} \end{cases} \tag{6}$$

*D. Address*

The attribute set = {<address>, <line1>, <line2>, <city>, <zipcode>, <state>, <province>, <country>}

Steps for measuring correlation of address:

1. If all the documents in the input set have child attributes,
   - 1.1. If all documents contain a <country>,
     A candidate <country> value $(\mathrm{Candidate_{country}})$ is selected based on majority voting. The selection process involves textual comparison and country code comparison
     - 1.1.1. $\mathrm{COUNTRY}_{\mathrm{d_k}} \neq \mathrm{Candidate_{country}}$
       - 1.1.1.1. $\mathrm{CS}_{\mathrm{d_k,address}} = 0$
   - 1.2. If <country> matched and if all documents contain <province>,
     Repeat Step 1.1 for <province>
   - 1.3. Repeat Step 1.2 for <state>, <zipcode> and <city>.

2. If at least one document contains the value of the address represented as a single value with no child attributes,
   - 2.1. The address values of the remaining documents are concatenated based on the order of the values represented in the document.
   - 2.2. Correlation score for the address is calculated using the method in Section IV Part A, from Step 2 onwards.

*E. National Identification Code*

The attribute set = {<nic>}

The similarity national identification code was measured by checking for exact matching values. If different values were obtained, the candidate method was used as in Section IV Part B.

$$\mathrm{CS}_{\mathrm{d_k,nic}} = \begin{cases} 1, & \mathrm{NIC}_{\mathrm{d_k}} = \mathrm{Candidate_{nic}} \\ 0, & \text{otherwise} \end{cases} \tag{7}$$

*F. Correlation score of a document*

The final correlation score is the average of the scores of all attributes.

The final correlation score of a document $\mathrm{d_k}$:

$$CS_{d_k} = \frac{1}{n_A}\sum_{i}^{n_A} CS_{d_k,A} \tag{8}$$

where $A$ is the attribute and $n_A$ is the number of attributes available.

## V. Conclusion and Future Work

The need of static and automated correlation calculation is gradually increasing in a world where data processes are rapidly replacing the human intervention with automated computations. Correlation between personal identity documents is typically measured by a human by comparing and analyzing a set of documents and identifying whether they belong to the same individual or not. In this paper, we have proposed and successfully implemented a solution that automates the correlation calculation and provides a normalized score that determines the correlation of a personal identity document within a set of documents.

The algorithm to calculate the correlation score of "name" and "address" attributes $(A)$ in a document $\mathrm{d_k}$ is:

$$CS_{d_k,A} = \left(\frac{1}{n-1}\right)\sum_{\substack{i=0 \\ i\neq k}}^{n} CS_{d_k,d_i,A} \tag{9}$$

The algorithm to calculate the correlation score of "date of birth", "gender" and "nic" attributes $(A)$ in a document $\mathrm{d_k}$ is:

$$\mathrm{CS}_{\mathrm{d_k,A}} = \begin{cases} 1, & \mathrm{Candidate_A} \\ 0, & \text{otherwise} \end{cases} \tag{10}$$

The overall correlation score of a document $d_k$ is the average of the attribute scores:

$$CS_{d_k} = \frac{1}{n_A} \sum_{i}^{n_A} CS_{d_k,A} \qquad (11)$$

where $A$ is the attribute and $n_A$ is the number of attributes available.

The algorithm is successfully implemented in IDStack, the common protocol for document verification built of digital signatures [22]. IDStack is a protocol that facilitates three modules: data extraction, data validation and scoring. Our algorithm is used in the score module which is used by a relying party to evaluate a set of documents sent by the owner of the documents.

The future contributions that can take this algorithm to the next level include fine-tuning the phonetic representation code and adding language extensions. We currently use Soundex to retrieve the phonetic representation of name segments and address segments and the traditional Soundex fail to differentiate names that end in differently pronounced vowels (Eg. Both Kasun and Kasuni have the same Soundex code even though the two names are pronounced differently). The current algorithm supports only the documents in English language which can be improved in the future to support multiple languages.